\documentclass[showkeys,showpacs,superscriptaddress]{revtex4}
\usepackage{amsmath}
\usepackage{amssymb}
\usepackage{graphicx}
\usepackage{color,ulem}
\allowdisplaybreaks

\begin{document}

\title{Energy conditions for a  $T^2$ wormhole at the center
}

\author{
Vladimir Dzhunushaliev
}
\email{v.dzhunushaliev@gmail.com}
\affiliation{
Department of Theoretical and Nuclear Physics,  Al-Farabi Kazakh National University, Almaty 050040, Kazakhstan
}
\affiliation{
Institute of Experimental and Theoretical Physics,  Al-Farabi Kazakh National University, Almaty 050040, Kazakhstan
}
\affiliation{
Institute of Physicotechnical Problems and Material Science of the NAS of the Kyrgyz Republic, 265 a, Chui Street, Bishkek 720071, Kyrgyzstan
}

\author{Vladimir Folomeev}
\email{vfolomeev@mail.ru}
\affiliation{
Institute of Experimental and Theoretical Physics,  Al-Farabi Kazakh National University, Almaty 050040, Kazakhstan
}
\affiliation{
Institute of Physicotechnical Problems and Material Science of the NAS of the Kyrgyz Republic, 265 a, Chui Street, Bishkek 720071, Kyrgyzstan
}

\author{
Burkhard Kleihaus
}
\email{
b.kleihaus@uni-oldenburg.de}
\affiliation{
Institut f\"ur Physik, Universit\"at Oldenburg, Postfach 2503
D-26111 Oldenburg, Germany
}

\author{
Jutta Kunz
}
\email{
jutta.kunz@uni-oldenburg.de
}
\affiliation{
Institut f\"ur Physik, Universit\"at Oldenburg, Postfach 2503
D-26111 Oldenburg, Germany
}

\date{\today}

\begin{abstract}
Within general relativity, we determine the energy conditions needed for the existence of a toroidal $T^2$ wormhole. For this purpose, we employ the conditions of the positiveness of the second derivatives of the relevant components of the metric, which describe an increase in the linear sizes (or the area) of the cross section of the throat. The corresponding inequalities for
the central energy density and pressures of the matter and for the metric are obtained.
\end{abstract}

\pacs{
}

\date{\today}

\maketitle

\section{Introduction}

The study of wormholes has a long history in General Relativity and in generalized theories of gravity (see, e.g., \cite{Visser:1995cc,Lobo:2017oab}).
The non-trival topology of wormhole solutions, where different regions of spacetime are connected via a throat, requires the presence of exotic matter in General Relativity
\cite{Ellis:1973yv,Bronnikov:1973fh,Kodama:1978dw,Ellis:1979bh,Morris:1988cz,Morris:1988tu,Lobo:2005us,Lobo:2017oab}, while in generalized theories of gravity the gravitational interaction itself may provide effective stress-energy tensors, that allow for the violation of the energy conditions
\cite{Hochberg:1990is,Fukutaka:1989zb,Ghoroku:1992tz,Furey:2004rq,Lobo:2009ip,Bronnikov:2009az,Kanti:2011jz,Kanti:2011yv,Harko:2013yb}.

Most previous studies of wormholes have considered throats of spherical topology. In the case of static spherically symmetric wormholes, such a throat is simply given by a sphere and represents a minimal surface of the spacetime.
When the throat is rotating, its geometry changes, since it becomes deformed due to the rotation
\cite{Kashargin:2007mm,Kashargin:2008pk,Kleihaus:2014dla,Chew:2016epf},
while its topology remains unchanged.

It appears interesting to consider also other throat topologies
\cite{Visser:1995cc,Lobo:2017oab}. For instance, cylindrical wormholes have been investigated in \cite{Bronnikov:2009na,Bronnikov:2013zxa,Bronnikov:2018uje},
which possess a topology $S^1\times I$. (Here the $S^1$ represents a circle in the $(x,y)$ plane, while $I$ corresponds to an interval on the $z$ axis.)
However, a particularly attractive throat topology is represented by a torus $T^2 = S^1 \times S^1$. But so far such toroidal wormholes have been addressed
only briefly in the literature
\cite{GonzalezDiaz:1996sr,Dzhunushaliev:2019qze}.

Within General Relativity the derivation of solutions describing a toroidal $T^2$ wormhole is an extremely complicated problem. The point is that the equations describing such a wormhole are systems of partial differential equations. This can already be seen from the fact that in toroidal coordinates
[see Eq.~\eqref{tor_metric}] the flat Minkowski spacetime metric depends on two coordinates. To obtain such solutions, it is necessary (a) to assign boundary conditions at the throat and at infinity; (b) to determine the properties of the matter needed to construct a $T^2$ wormhole; (c) and finally to obtain solutions of the partial differential equations (the Einstein-matter equations) subject to these boundary conditions. Here asymptotically flat solutions are evidently of most interest.

Consequently, we expect that the problem of obtaining solutions describing toroidal $T^2$ wormholes should be split into several stages: (a) studying the properties of the matter needed to obtain toroidal wormholes; (b) investigating the asymptotic behavior of the solutions for $T^2$ wormholes; and (c) obtaining and solving the set of differential equations describing such wormholes subject to the appropriate boundary conditions. Presumably these solutions should be sought numerically.

The present paper is a continuation of the study performed in Ref.~\cite{Dzhunushaliev:2019qze}, where we have obtained and studied a toroidal thin-shell wormhole. Here, we analyze the conditions imposed on the matter needed for the existence of a toroidal  $T^2$ wormhole. To do this, we write down the Einstein-matter equations at the throat and assign the necessary geometric conditions providing the existence of a throat. For these conditions we take the condition of the positiveness of the second derivatives of the relevant components of the metric, which describe an increase in the linear sizes (or the area) of the cross section of the throat.

\section{Energy conditions for a  $S^2$ wormhole at the throat}
\label{S2_WH}

To begin with, let us recall the procedure of obtaining the energy conditions for a $S^2$ wormhole at the throat, in order to repeat it for a toroidal $T^2$ wormhole in the next section. Let us take the following metric for a $S^2$ wormhole:
\begin{equation}
	ds^2 = A(r) dt^2 - dr^2 - B(r) \left(
		d \theta^2 + \sin^2 \theta d \varphi^2
	\right) .
\label{1_10}
\end{equation}
The Einstein equations are
\begin{equation}
	R^\mu_\nu - \frac{1}{2} \delta^\mu_\nu R = \varkappa T^\mu_\nu,
\label{1_20}
\end{equation}
where $\varkappa=8\pi G$,
and the energy-momentum tensor for the macroscopic matter is taken in the form
\begin{equation}
	T^\mu_\nu = \text{diag}\left(\epsilon, -p_r, -p_\theta, -p_\varphi\right).
\label{1_30}
\end{equation}
In order for this energy-momentum tensor to be consistent with the spherically symmetric metric \eqref{1_10}, it is necessary to take $ p_\varphi= p_\theta$.
Then the Einstein equations \eqref{1_20} with the metric \eqref{1_10} yield the following equations:
\begin{eqnarray}
	\frac{A^{\prime \prime}}{A} + \frac{A^\prime B^\prime}{A B} -
	\frac{{A^\prime}^2}{2 A^2} &=& \varkappa \left(
		\epsilon + p_r + 2 p_\theta
	\right) ,
\label{1_40}\\
	\frac{B^{\prime \prime}}{B} - \frac{A^\prime B^\prime}{2 A B} -
	\frac{{B^\prime}^2}{2 B^2} &=& -\varkappa \left(
		 \epsilon + p_r
	\right) .
\label{1_50}
\end{eqnarray}
Bearing in mind that the metric functions $A(r)$ and $B(r)$ must be even,
we have the following expressions for the second derivatives at the throat:
\begin{eqnarray}
	\frac{A^{\prime \prime}_0}{A_0} &=& \varkappa \left[
		\epsilon_0 + \left( p_r\right)_0 + 2\left( p_\theta\right)_0
	\right] ,
\label{1_60}\\
	\frac{B^{\prime \prime}_0}{B_0} &=& -\varkappa \left[
		 \epsilon_0 + \left( p_r\right)_0
	\right].
\label{1_70}
\end{eqnarray}
Here the index 0 indicates that the value of a quantity is taken at the throat.
For a spherically symmetric wormhole the components of the metric
$g_{\theta \theta} = -B$ and $g_{\varphi \varphi} = - B \sin^2 \theta$
should have a minimum at the throat. This gives the known relation for the ``exotic'' matter supporting the wormhole
\begin{equation}
	\epsilon_0 + \left( p_r\right)_0 < 0 .
\label{1_90}
\end{equation}

\section{Energy conditions for a  $T^2$ wormhole at the throat}

For a toroidal $T^2$ wormhole, we use the following metric:
\begin{equation}
	ds^2 = f(\chi, \beta) d t^2 - l(\chi, \beta) d \chi^2 -
	g(\chi, \beta) d \beta^2 - \omega(\chi, \beta) d \varphi^2.
\label{2_10}
\end{equation}
Here  $t, \chi, \beta, \varphi$ are toroidal coordinates which,
in a flat spacetime, describe the Minkowski metric as follows
\begin{equation}
	ds^2 = dt^2 - \left(  \frac{a}{\cosh \alpha - \cos \beta} \right)^2
	\left(
		d \alpha^2 + d \beta^2 + \sinh^2 \alpha d \varphi^2
	\right),
\label{tor_metric}
\end{equation}
where $a$ is some parameter and the coordinate  $\chi$ in \eqref{2_10}
is related to the coordinate $\alpha$ from  \eqref{tor_metric} as
$\alpha = - \ln \chi$. Using the Einstein equations \eqref{1_20}
and choosing the energy-momentum tensor as
\begin{equation}
	T^{\mu}_{\phantom{\mu} \nu} = \begin{pmatrix}
		\epsilon	&	0	&	0	&	0 \\
		0	&	-p_\chi	&	\frac{\sigma}{l}	&	0	\\
		0	&	\frac{\sigma}{g}	&	-p_\beta	&	0	\\
		0	&	0	&	0	&	-p_\varphi
	\end{pmatrix},
\label{2_15}
\end{equation}
we have the following equations for a toroidal  $T^2$ wormhole:
\begin{eqnarray}
	&& - \frac{\omega_{,\chi, \chi}}{\omega} -
	\frac{l}{g}\frac{\omega_{,\beta, \beta}}{\omega} -
	\frac{g_{,\chi, \chi}}{g} - \frac{l_{,\beta, \beta}}{g} +
	\frac{g_{,\beta} l_{,\beta}}{2 g^2} +
	\frac{l}{2 }\frac{g_{,\beta} \omega_{,\beta}}{g^2 \omega} +
	\frac{g^2_{,\chi}}{2 g^2} +
	\frac{g_{,\chi} l_{,\chi}}{2 l g} -
	\frac{g_{,\chi} \omega_{,\chi}}{2 g \omega} +
	\frac{l^2_{,\beta}}{2 l g} -
	\frac{l_{,\beta} \omega_{,\beta}}{2 g \omega}
\nonumber \\
	&&
	+\frac{l_{,\chi} \omega_{,\chi}}{2 l \omega} +
	\frac{l}{2 g}\frac{\omega^2_{,\beta}}{\omega^2} +
	\frac{\omega^2_{,\chi}}{2 \omega^2}
	= 2 \varkappa \,l\, \epsilon ,
\label{2_20}\\
	&& - \frac{\omega_{,\beta, \beta}}{\omega} -
	\frac{f_{,\beta, \beta}}{f} +
	\frac{g_{,\beta} f_{,\beta}}{2 g f} +
	\frac{g_{,\beta} \omega_{,\beta}}{2 g \omega}  -
	\frac{g_{,\chi} f_{,\chi}}{2 l f} -
	\frac{g_{,\chi} \omega_{,\chi}}{2 l \omega} +
	\frac{f^2_{,\beta}}{2 f^2} -
	\frac{f_{,\beta} \omega_{,\beta}}{2 f \omega} -
	\frac{g}{2 l}\frac{f_{,\chi} \omega_{,\chi}}{f \omega} +
	\frac{\omega^2_{,\beta}}{2 \omega^2}
	= - 2 \varkappa \, g\, p_\chi ,
\label{2_30}\\
	&&
	\frac{f_{,\beta ,\chi}}{f} + \frac{\omega_{,\beta ,\chi}}{\omega} -
	\frac{g_{,\chi} f_{,\beta}}{2 g f} -
	\frac{g_{,\chi} \omega_{,\beta}}{2 g \omega} -
	\frac{l_{,\beta} f_{,\chi}}{2 l f} -
	\frac{l_{,\beta} \omega_{,\chi}}{2 l \omega} -
	\frac{f_{,\chi} f_{,\beta}}{2 f^2} -
	\frac{\omega_{,\chi} \omega_{,\beta}}{2 \omega^2} =
	2 \varkappa \sigma ,
\label{2_40}\\
	&&
	- \frac{f_{,\chi, \chi}}{f} - \frac{\omega_{,\chi, \chi}}{\omega} -
	\frac{l_{,\beta} f_{,\beta}}{2 g f} -
	\frac{l_{,\beta} \omega_{,\beta}}{2 g \omega}  -
	\frac{l}{2 g}\frac{f_{,\beta} \omega_{,\beta}}{f \omega} +
	\frac{l_{,\chi} f_{,\chi}}{2 l f} +
	\frac{l_{,\chi} \omega_{,\chi}}{2 l \omega} +
	\frac{f^2_{,\chi}}{2 f^2} -
	\frac{f_{,\chi} \omega_{,\chi}}{2 f \omega} +
	\frac{\omega^2_{,\chi}}{2 \omega^2} =
	-2 \varkappa \,l\, p_\beta ,
\label{2_50}\\
	&&
	- \frac{f_{,\chi ,\chi}}{f} - \frac{g_{,\chi ,\chi}}{g} -
	\frac{l}{g}\frac{f_{,\beta ,\beta}}{f} -
	\frac{l_{,\beta ,\beta}}{g} +
	\frac{g_{,\beta} l_{,\beta}}{2 g^2} +
	\frac{l}{2 }\frac{g_{,\beta} f_{,\beta}}{g f} +
	\frac{g^2_{,\chi}}{2 g^2} +
	\frac{g_{,\chi} l_{,\chi}}{2 g l} -
	\frac{g_{,\chi} f_{,\chi}}{2 g f} +
	\frac{l^2_{,\beta}}{2l g} -
	\frac{l_{,\beta} f_{,\beta}}{2 g f}
\nonumber \\
	&&
	+\frac{l}{2 g} \frac{f^2_{,\beta}}{f^2} +
	\frac{l_{,\chi} f_{,\chi}}{2 l f} +
	\frac{f_{,\chi}^2}{2 f^2} =
	- 2 \varkappa \,l\, p_\varphi .
\label{2_60}
\end{eqnarray}
To determine the energy conditions imposed on the energy-momentum tensor
of the matter supporting the wormhole, we, analogously to Sec.~\ref{S2_WH},
write down the
$
\left( ^t_t \right), \left( ^\beta_\beta \right)$, and
$\left( ^\varphi_\varphi \right)$
components of the Einstein equations at the throat, i.e., at  $\chi = 0$,
solving them with respect to higher-order derivatives
$f_{,\chi, \chi}, g_{,\chi, \chi}$, and $\omega_{,\chi, \chi}$:
\begin{eqnarray}
	\left( f_{, \chi ,\chi} \right) _0 &=& \varkappa f_0 l_0 \left[
		\epsilon_0 + \left( p_\beta \right)_0 + \left( p_\varphi \right)_0
	\right] + \frac{f_0 l_0}{2 \omega_0 g_0} \left( \omega_{,\beta ,\beta} \right)_0 -
	\frac{f_0 l_0}{4 g_0} \frac{\left( \omega^2_{, \beta}\right)_0}{\omega^2_0} -
	\frac{f_0 l_0}{4 g_0}
	\frac{\left( \omega_{, \beta}\right)_0 \left( g_{, \beta}\right)_0}
	{\omega_0 g_0} -
	l_0 \frac{\left(f_{, \beta}\right)_0 \left(\omega_{, \beta}\right)_0}
	{4 \omega_0 g_0}
\nonumber \\
	&&
	+l_0 \frac{\left(g_{, \beta}\right)_0 \left(f_{, \beta}\right)_0}{4 g^2_0} -
	\frac{\left(l_{, \beta}\right)_0 \left( f_{, \beta}\right)_0}{2 g_0} -
	l_0 \frac{\left( f_{, \beta ,\beta}\right)_0}{2 g_0} +
	l_0 \frac{\left( f_{, \beta}^2\right)_0}{4 f_0 g_0} ,
\label{2_70}\\
	\left( g_{, \chi ,\chi} \right) _0 &=& \varkappa g_0 l_0 \left[
			- \epsilon_0 - \left( p_\beta \right)_0 + \left( p_\varphi \right)_0
		\right] -
	\left( l_{,\beta ,\beta} \right)_0 -
	l_0 \frac{\left(\omega_{, \beta ,\beta}\right)_0}{2 \omega_0} +
	l_0 \frac{\left(\omega^2_{, \beta}\right)_0}{4 \omega_0^2} +
	l_0 \frac{\left(\omega_{, \beta}\right)_0 \left(g_{, \beta}\right)_0}
	{4 g_0 \omega_0} +
	\frac{\left(g_{, \beta}\right)_0 \left(l_{, \beta}\right)_0}{2 g_0}
\nonumber \\
	&&
	+\frac{\left(l_{, \beta}^2\right)_0}{2 l_0} +
	l_0 \frac{\left(f_{, \beta}\right)_0 \left(\omega_{, \beta}\right)_0}
	{4 \omega_0 f_0} +
	l_0 \frac{\left(g_{, \beta}\right)_0 \left(f_{, \beta}\right)_0}{4 g_0 f_0} -
	l_0 \frac{\left(f_{, \beta ,\beta}\right)_0}{2 f_0} +
	l_0 \frac{\left(f_{, \beta}^2\right)_0}{4 f_0^2} ,
\label{2_80}\\
	\left( \omega_{, \chi ,\chi} \right) _0 &=& \varkappa \omega_0 l_0 \left[
		- \epsilon_0 + \left( p_\beta \right)_0 - \left( p_\varphi \right)_0
	\right] - \frac{l_0}{2 g_0} \left( \omega_{, \beta ,\beta} \right)_0 +
	\frac{l_0}{4 g_0 \omega_0} \left( \omega^2_{,\beta}\right)_0 +
	\frac{l_0}{4 g_0^2}
	\left( \omega_{, \beta}\right)_0 \left( g_{, \beta}\right)_0 -
	\frac{\left(\omega_{, \beta}\right)_0 \left(l_{, \beta}\right)_0}{2 g_0}
\nonumber \\
	&&
	-\frac{l_0}{4 f_0 g_0} \left(f_{, \beta}\right)_0
	\left(\omega_{, \beta}\right)_0 -
	\frac{\omega_0 l_0}{4 f_0 g_0^2}
	\left(g_{, \beta}\right)_0 \left( f_{, \beta} \right)_0 +
	\frac{\omega_0 l_0}{2 f_0 g_0} \left(f_{, \beta ,\beta}\right)_0 -
	\frac{\omega_0 l_0}{4 f_0^2 g_0} \left( f_{, \beta}^2\right)_0 .
\label{2_90}
\end{eqnarray}
Here we have taken into account that all functions are even, 
\begin{equation*}
	\left.
	\frac{\partial}{\partial \chi} 
	\biggl[
		f(\chi, \beta), l(\chi, \beta), g(\chi, \beta), \omega(\chi, \beta)
		\biggl]
	\right|_{\chi = 0} = 0,
\end{equation*}
i.e., the wormhole is symmetric. 
The $\left( ^\chi_{\chi} \right)$ component of the Einstein equations is
\begin{equation}
	\frac{1}{2 g_0} \left[
		\frac{\left( f_{, \beta ,\beta}\right)_0}{f_0} +
		\frac{\left( \omega_{, \beta ,\beta}\right)_0}{\omega_0} -
		\frac{\left(\omega^2_{, \beta}\right)_0}{2 \omega_0^2} -
		\frac{\left(\omega_{, \beta}\right)_0 \left(g_{, \beta}\right)_0}
		{2 g_0 \omega_0} +
		\frac{\left(\omega_{, \beta}\right)_0 \left(f_{, \beta}\right)_0}
		{2\omega_0 f_0} -
		\frac{\left(g_{, \beta}\right)_0 \left(f_{, \beta}\right)_0}{2 g_0 f_0} -
		\frac{\left(f^2_{, \beta}\right)_0}{2 f_0^2}
	\right] = \varkappa \left( p_\chi \right)_0 .
\label{2_100}
\end{equation}
The component $\left( ^\chi_\beta \right)$ and the equation
\begin{equation}
	T^\mu_{\phantom{\mu} \nu ; \mu} = 0
\label{2_110}
\end{equation}
for $\nu = \chi$ are satisfied when
\begin{equation}
	\sigma_0(\beta) = \sigma(\chi = 0, \beta) = 0 .
\label{2_120}
\end{equation}
Eq.~\eqref{2_110} is satisfied for $\nu = t, \varphi$, and when
 $\nu = \beta$ it has the following form:
 \begin{equation}
	\frac{\left( l_{, \beta}\right)_0}{l_0} \left( p_\chi \right)_0 +
	\left[
		\frac{\left( f_{, \beta}\right)_0}{f_0} -
		\frac{\left( l_{, \beta}\right)_0}{l_0} -
		\frac{\left( \omega_{, \beta}\right)_0}{\omega_0}
	\right] \left( p_\beta \right)_0 -
	2 \left( p_{\chi , \beta} \right)_0 -
	\frac{\left( f_{, \beta}\right)_0}{f_0} \epsilon_0 +
	\frac{\left( \omega_{, \beta}\right)_0}{\omega_0} \left( p_\varphi \right)_0
	= 0 .
\label{2_130}
\end{equation}
As usual, we assume that the necessary condition for a wormhole to exist
(in the present case, a $T^2$  wormhole)
is the presence of minima of the metric functions
$g_{\beta \beta}$ and $g_{\varphi \varphi}$:
\begin{equation}
	\left.
		\frac{\partial^2 (g, \omega)}{\partial \chi^2}
	\right|_{\chi = 0} > 0 .
\label{2_150}
\end{equation}
Note that, instead of the conditions \eqref{2_150}, in Refs.~\cite{Bronnikov:2009na,Bronnikov:2013zxa}, the authors discuss a necessary condition for the existence of a wormhole according to which a throat has a minimum of its 2-dimensional space cross section. In our case this corresponds to a minimum of the product $g \times \omega$,
to yield
\begin{equation}
	\left.
		\left( \frac{1}{g} \frac{\partial^2 g}{\partial \chi^2} \right)
	\right|_{\chi = 0} +
	\left.
		\left( \frac{1}{\omega} \frac{\partial^2 \omega}{\partial \chi^2} \right)
	\right|_{\chi = 0} > 0 .
\label{2_160}
\end{equation}
Here we have taken into account the condition that the area of throat has a minimum, i.e., $\partial \left( g \omega\right)/\partial \chi = 0$.

Here we will consider the conditions \eqref{2_150}, which are more strict than \eqref{2_160}. This ensures that the cross section of a  $T^2$ wormhole increases along both radii, when one moves away from the throat. In other words, the lengths of the circles  $2 \pi \sqrt g$ and $2 \pi \sqrt \omega$ will increase, when one moves away from the throat. On the other hand, in the case when the condition \eqref{2_160} is satisfied, the area of the cross section will increase, but the length of one of the circles may decrease while the length of the other one increases.

Thus Eq.~\eqref{2_150} yields the following conditions imposed on the energy density and pressures of the matter needed to create a toroidal $T^2$ wormhole
[they follow from Eqs.~\eqref{2_80} and \eqref{2_90}, respectively]:
\begin{eqnarray}
	&&
	\varkappa \left[
			\epsilon_0 + \left( p_\beta \right)_0 - \left( p_\varphi \right)_0
		\right] < \frac{1}{2 g_0} \left\{
		\left[
			-	\frac{\left(\omega_{, \beta ,\beta}\right)_0}{\omega_0} +
			\frac{\left(\omega^2_{, \beta}\right)_0}{2 \omega_0^2} +
			\frac{\left(\omega_{, \beta}\right)_0 \left(g_{, \beta}\right)_0}
			{2 g_0 \omega_0}
 		\right]
	\right.
\nonumber \\
	&&
	+\left.
		\left[
			-	\frac{\left(f_{, \beta ,\beta}\right)_0}{f_0} +
			\frac{\left(f_{, \beta}^2\right)_0}{2 f_0^2} +
			\frac{\left(f_{, \beta}\right)_0 \left(\omega_{, \beta}\right)_0}
			{2 \omega_0 f_0} +
			\frac{\left(g_{, \beta}\right)_0 \left(f_{, \beta}\right)_0}
			{2 g_0 f_0}
		\right] -
		2 \frac{\left( l_{,\beta ,\beta} \right)_0}{l_0} +
 		\frac{\left(g_{, \beta}\right)_0 \left(l_{, \beta}\right)_0}{g_0 l_0} +
		\frac{\left(l_{, \beta}^2\right)_0}{l_0^2}
	\right\} ,
\label{2_180}\\
	&&
	\varkappa \left[
		\epsilon_0 - \left( p_\beta \right)_0 + \left( p_\varphi \right)_0
		\right] < \frac{1}{2 g_0} \left\{
		\left[
			-	\frac{\left(\omega_{, \beta ,\beta}\right)_0}{\omega_0} +
			\frac{\left(\omega^2_{, \beta}\right)_0}{2 \omega_0^2} +
			\frac{\left(\omega_{, \beta}\right)_0 \left(g_{, \beta}\right)_0}
			{2 g_0 \omega_0}
 		\right]
	\right.
\nonumber \\
	&&
	-\left.
	\left[
		-	\frac{\left(f_{, \beta ,\beta}\right)_0}{f_0} +
		\frac{\left(f_{, \beta}^2\right)_0}{2 f_0^2} +
		\frac{\left(f_{, \beta}\right)_0 \left(\omega_{, \beta}\right)_0}
		{2 \omega_0 f_0} +
		\frac{\left(g_{, \beta}\right)_0 \left(f_{, \beta}\right)_0}
		{2 g_0 f_0}
	\right] -
	\frac{\left(\omega_{, \beta}\right)_0 \left( l_{, \beta} \right)_0}
	{\omega_0 l_0}
	\right\} .
\label{2_190}
\end{eqnarray}
Taking into account Eq.~\eqref{2_100}, the inequalities~\eqref{2_180}
and \eqref{2_190} can be rewritten in a simpler form to give
\begin{eqnarray}
	\varkappa \left[
		\epsilon_0 + \left( p_\chi \right)_0 + \left( p_\beta \right)_0 -
		\left( p_\varphi \right)_0
		\right] &<& \frac{1}{g_0} \left[
		- \frac{\left( l_{,\beta ,\beta} \right)_0}{l_0} +
		\frac{\left(l_{, \beta}^2\right)_0}{2 l_0^2} +
		\frac{\left(g_{, \beta}\right)_0 \left(l_{, \beta}\right)_0}{2 g_0 l_0} +
		\frac{\left(f_{, \beta}\right)_0 \left(\omega_{, \beta}\right)_0}
		{2 \omega_0 f_0}
	\right] ,	
\label{2_210}\\
	\varkappa \left[
		\epsilon_0 + \left( p_\chi \right)_0 - \left( p_\beta \right)_0 +
		\left( p_\varphi \right)_0
		\right] &<& \frac{1}{g_0} \left[
		\frac{\left(f_{, \beta ,\beta}\right)_0}{f_0} -
		\frac{\left( f_{, \beta}^2\right)_0}{2 f_0^2}	-
		\frac{\left(g_{, \beta}\right)_0 \left( f_{, \beta} \right)_0}
		{2 g_0 f_0} -
		\frac{\left(\omega_{, \beta}\right)_0 \left(l_{, \beta}\right)_0}
		{2 \omega_0 l_0}
	\right] .
\label{2_220}
\end{eqnarray}
The inequality~\eqref{2_160} describing a minimum of the area of the throat
can be rewritten in the form
\begin{equation}
\begin{split}
	\varkappa \epsilon_0 < & \frac{1}{2 g_0} \left[
		- \frac{\left( l_{,\beta ,\beta} \right)_0}{l_0}
		- \frac{\left( \omega_{, \beta ,\beta} \right)_0}{\omega_0} +
		\frac{\left( \omega^2_{, \beta}\right)_0}{2 \omega^2_0} +
		\frac{\left(l_{, \beta}^2\right)_0}{2 l_0^2} +
		\frac{\left( \omega_{, \beta}\right)_0 \left( g_{, \beta}\right)_0}
		{2 \omega_0 g_0} -
		\frac{\left(\omega_{, \beta}\right)_0 \left(l_{, \beta}\right)_0}
		{2 \omega_0 l_0} +
		\frac{\left(g_{, \beta}\right)_0 \left(l_{, \beta}\right)_0}{2 g_0 l_0}
	\right].
\end{split}
\label{2_230}
\end{equation}

\section{Analysis of the energy conditions for a  $T^2$ wormhole}

In constructing wormhole solutions, it is of great interest to study the question of violation of the energy conditions at the throat: whether such a violation is necessary for the throat to exist? For a static $S^2$ wormhole, the answer is positive. In this section we consider some particular conditions
of violation (or nonviolation) of the energy conditions for a $T^2$ wormhole.

For convenience of performing calculations, let us introduce new functions
\begin{equation}
	f(\chi, \beta) = e^{F(\chi, \beta)} , \quad	
	g(\chi, \beta) = e^{G(\chi, \beta)} , \quad
	l(\chi, \beta) = e^{L(\chi, \beta)} , \quad
	\omega(\chi, \beta) = e^{\Omega(\chi, \beta)} .
\label{3_1_10}
\end{equation}
Using them, we analyze the conditions \eqref{2_180} and \eqref{2_190}
which are necessary for the existence of minima of the metric components
$g_{\beta \beta}$ and $g_{\varphi \varphi}$.
The inequalities \eqref{2_180} and \eqref{2_190} can be reduced
to a more symmetric form if we set
  \begin{equation}
	L_{,\beta}(\chi = 0, \beta) = 0 .
\label{3_2_10}
\end{equation}
Then the first terms on the left- and right-hand sides of the inequalities \eqref{2_180} and \eqref{2_190} are the same, and the second terms have different signs.

To satisfy these inequalities, one can consider the following particular case
when the second terms on the left-hand sides are equal
to the second terms on the right-hand sides:
\begin{equation}
	\varkappa \left[
		\left( p_\beta \right)_0 - \left( p_\varphi \right)_0
	\right] = \frac{e^{-G}}{2}
	\left[
		-	\left(F_{, \beta ,\beta}\right)_0 -
		\frac{\left(F_{, \beta}^2\right)_0}{2} +
		\frac{\left(F_{, \beta}\right)_0 \left(\Omega_{, \beta}\right)_0}{2} +
		\frac{\left(F_{, \beta}\right)_0 \left(G_{, \beta}\right)_0}{2}
	\right] .
\label{3_2_20}
\end{equation}
In the following we will study the consequences for this particular case. 
The inequalities \eqref{2_180} and \eqref{2_190} are now identical and read
\begin{equation}
	\varkappa \epsilon_0  < \frac{e^{-G}}{2} \left[
		-	\left(\Omega_{, \beta ,\beta}\right)_0 -
		\frac{\left(\Omega^2_{, \beta}\right)_0}{2} +
		\frac{\left(\Omega_{, \beta}\right)_0 \left(G_{, \beta}\right)_0}{2}
	\right] ,
\label{3_2_30}
\end{equation}
Thus we have Eqs. \eqref{2_100}, \eqref{2_130}, \eqref{3_2_20} and inequality \eqref{3_2_30}.

Eqs.~\eqref{2_100}, \eqref{2_130} and \eqref{3_2_20} can be solved
with respect to the pressures,
$
\left( p_\chi \right)_0, \left( p_\beta \right)_0$, and $\left( p_\varphi \right)_0
$,
\begin{eqnarray}
	\left( p_\chi \right)_0 &=& \frac{e^{-G_0}}{2 \varkappa} \left[
		\left( F_{, \beta, \beta}\right)_0 +
		\frac{\left( F_{, \beta}^2\right)_0}{2} +
		\left( \Omega_{, \beta, \beta}\right)_0 +
		\frac{\left( \Omega_{, \beta}^2\right)_0}{2} +
		\frac{\left(F_{, \beta}\right)_0 \left(\Omega_{, \beta}\right)_0}{2} -
		\frac{\left(F_{, \beta}\right)_0 \left( G_{, \beta}\right)_0}{2} -
		\frac{\left(G_{, \beta}\right)_0 \left(\Omega_{, \beta}\right)_0}{2}
	\right] ,
\label{3_2_60} \\
	\left( p_\beta \right)_0 &=& \epsilon_0 +
	2 \frac{\left( p_{\chi , \beta} \right)_0}{F_{, \beta}} +
	\frac{e^{-G_0}}{2 \varkappa} \left\{
		- \frac{\left(F_{,\beta, \beta}\right)_0\left(\Omega_{,\beta}\right)_0}
		{\left(F_{, \beta}\right)_0} +
		\frac{\left(\Omega_{, \beta}\right)^2_0}{2} +
		\frac{\left(\Omega_{, \beta}\right)_0}{2} \left[
			- \left(F_{, \beta}\right)_0 + \left(G_{, \beta}\right)_0
		\right]
	\right\} ,
\label{3_2_70}\\
	\left( p_\varphi \right)_0 &=& \epsilon_0 +
		2 \frac{\left( p_{\chi , \beta} \right)_0}{F_{, \beta}} +
	\frac{e^{-G_0}}{2 \varkappa} \left\{
	\left(F_{,\beta, \beta}\right)_0 \left[
			1 - \frac{\left(\Omega_{,\beta}\right)_0}{\left(F_{, \beta}\right)_0}
		\right] +
		\frac{\left(F_{, \beta}\right)_0^2}{2} -
		\frac{\left(F_{, \beta}\right)_0 \left(G_{, \beta}\right)_0}{2} +
		\frac{\left(\Omega_{, \beta}\right)^2_0}{2}
	\right.
\nonumber \\
	&&
	+\left.
			\frac{\left(\Omega_{, \beta}\right)_0}{2} \left[
			- 2 \left(F_{, \beta}\right)_0 + \left(G_{, \beta}\right)_0
		\right]
	\right\} .
\label{3_2_80}
\end{eqnarray}
Let us now analyze the energy conditions.

\subsection{The null energy condition}

In general the null energy condition asserts that for any null vector $k_\mu$
\begin{equation}
	T_{\mu \nu} k^\mu k^\nu \geq 0 ,
\label{3_2_90}
\end{equation}
or in terms of the principal pressures $p_i$
\begin{equation}
	\epsilon_0 + p_i \geq 0, \quad i = \chi, \beta, \varphi.
\label{3_2_100}
\end{equation}
In our particular case Eq.~(\ref{3_2_20}) we have the following expressions for the left-hand sides of \eqref{3_2_100}:
\begin{eqnarray}
	\epsilon_0 + \left( p_\chi \right)_0 &=& \epsilon_0 +
	\frac{e^{-G_0}}{2 \varkappa} \left[
		\left( F_{, \beta, \beta}\right)_0 +
		\frac{\left( F_{, \beta}^2\right)_0}{2} +
		\left( \Omega_{, \beta, \beta}\right)_0 +
		\frac{\left( \Omega_{, \beta}^2\right)_0}{2} +
		\frac{\left(F_{, \beta}\right)_0 \left(\Omega_{, \beta}\right)_0}{2} -
		\frac{\left(F_{, \beta}\right)_0 \left( G_{, \beta}\right)_0}{2}
	\right.
\nonumber \\		
	&&
	-\left.
		\frac{\left(G_{, \beta}\right)_0 \left(\Omega_{, \beta}\right)_0}{2}
	\right] ,
\label{3_2_110}\\
	\epsilon_0 + \left( p_\beta \right)_0 &=& 2 \epsilon_0 +
	2 \frac{\left( p_{\chi , \beta} \right)_0}{F_{, \beta}} +
	\frac{e^{-G_0}}{2 \varkappa} \left\{
		- \frac{\left(F_{,\beta, \beta}\right)_0\left(\Omega_{,\beta}\right)_0}
		{\left(F_{, \beta}\right)_0} +
		\frac{\left(\Omega_{, \beta}\right)^2_0}{2} +
		\frac{\left(\Omega_{, \beta}\right)_0}{2} \left[
			- \left(F_{, \beta}\right)_0 + \left(G_{, \beta}\right)_0
		\right]
	\right\} ,
\label{3_2_120}\\
	\epsilon_0 + \left( p_\varphi \right)_0 &=& 2 \epsilon_0 +
		2 \frac{\left( p_{\chi , \beta} \right)_0}{F_{, \beta}} +
	\frac{e^{-G_0}}{2 \varkappa} \left\{
	\left(F_{,\beta, \beta}\right)_0 \left[
			1 - \frac{\left(\Omega_{,\beta}\right)_0}{\left(F_{, \beta}\right)_0}
		\right] +
		\frac{\left(F_{, \beta}\right)_0^2}{2} -
		\frac{\left(F_{, \beta}\right)_0 \left(G_{, \beta}\right)_0}{2} +
		\frac{\left(\Omega_{, \beta}\right)^2_0}{2}
	\right.
\nonumber \\
	&&
	+\left.
			\frac{\left(\Omega_{, \beta}\right)_0}{2} \left[
			- 2 \left(F_{, \beta}\right)_0 + \left(G_{, \beta}\right)_0
		\right]
	\right\} .
\label{3_2_130}
\end{eqnarray}

\subsection{The weak energy condition}

The weak energy condition asserts in general that for any timelike vector $V_\mu$
\begin{equation}
	T_{\mu \nu} V^\mu V^\nu \geq 0 .
\label{3_3_10}
\end{equation}
In our case this gives
\begin{eqnarray}
	\epsilon_0 &\geq& 0,
\label{3_3_20}\\
	\epsilon_0 + p_i &\geq& 0, \quad i = \chi, \beta, \varphi.
\label{3_3_30}
\end{eqnarray}
The energy density $\epsilon_0$ satisfies the inequality \eqref{3_2_30}
and it can be positive if the right-hand side of this inequality will be positive in a whole range $- \pi \leq \beta \leq \pi$. In the particular case Eq.~(\ref{3_2_20}) the left-hand sides of the inequality \eqref{3_3_30} have the form \eqref{3_2_110}-\eqref{3_2_130}.

\subsection{The strong energy condition}

The strong energy condition asserts in general that for any timelike vector $V_\mu$
\begin{equation}
	\left(
		T_{\mu \nu} - \frac{1}{2} g_{\mu \nu} T
	\right) V^\mu V^\nu \geq 0 .
\label{3_4_10}
\end{equation}
In our case we then have
\begin{eqnarray}
	\epsilon_0 + \left( p_{i} \right)_0 &\geq& 0 , \quad
	i = \chi, \beta, \varphi ,
\label{3_4_20}\\
	\epsilon_0 + \sum\limits_{i} \left( p_{i} \right)_0 &\geq& 0.
\label{3_4_30}
\end{eqnarray}
The energy density satisfies the inequality \eqref{3_2_30} and it can be positive if the right-hand side of this inequality will be positive in a whole range $- \pi \leq \beta \leq \pi$. In the particular case Eq.~(\ref{3_2_20}) and taking into account Eqs.~\eqref{3_2_110}-\eqref{3_2_130}, the left-hand side of the inequality \eqref{3_4_30} takes the form
\begin{equation}
\begin{split}
	\epsilon_0 + \left( p_{\chi} \right)_0 + \left( p_{\beta} \right)_0 +
	\left( p_{\varphi} \right)_0 = &
	3 \epsilon_0 + 4 \frac{\left( p_{\chi , \beta} \right)_0}{F_{, \beta}} +
	\frac{e^{-G_0}}{2 \varkappa} \left\{
		\left( \Omega_{, \beta, \beta}\right)_0 +
		2 \left(F_{,\beta, \beta}\right)_0 \left[
			1 - \frac{\left(\Omega_{,\beta}\right)_0}{\left(F_{, \beta}\right)_0}
		\right] + \frac{3 \left(\Omega_{, \beta}\right)^2_0}{2}
	\right.
\\
	&
	+\left.
		\frac{\left(\Omega_{, \beta}\right)_0 \left(G_{, \beta}\right)_0}{2} -
		\left(\Omega_{, \beta}\right)_0 \left(F_{, \beta}\right)_0 +
		\left(F_{, \beta}\right)_0^2 -
		\left(F_{, \beta}\right)_0 \left(G_{, \beta}\right)_0
	\right\rbrace.
\label{3_3_40}
\end{split}
\end{equation}
Summarizing, we see that to construct  a toroidal $T^2$ throat,
there are some necessary conditions to be imposed on the  matter supporting the wormhole. Of course, this does not ensure the existence of an asymptotically flat wormhole; to obtain such a wormhole, it is necessary to assign also asymptotic boundary conditions providing the asymptotic flatness of spacetime.

\section{Particular cases}

We see that even if $L_{,\beta} = 0$ the inequality \eqref{3_2_30} and the expressions for the pressures \eqref{3_2_60}-\eqref{3_2_80} are too cumbersome to perform the analysis of the conditions imposed on the matter, which are necessary for the existence of a toroidal $T^2$ wormhole. Therefore in this section we consider some particular cases permitting a simplification of the equations.

\subsection{Particular case with the positive right-hand side
of the inequality~\eqref{3_2_30}}

Consider the conditions providing the positiveness of the right-hand side of the inequality \eqref{3_2_30}. This assumes that the energy density
$\epsilon_0 > 0$. For this purpose, we take
\begin{eqnarray}
	\left( \Omega(\beta) \right)_0 &=& - \cos \beta ,
\label{5_1_10}\\
	\left( G(\beta) \right)_0 &=& - 3 \cos \beta +
	\log \sin^2 \beta .
\label{5_1_20}
\end{eqnarray}
In this case the inequality \eqref{3_2_30} yields
\begin{equation}
	\varkappa \epsilon_0 < \frac{e^{3 \cos \beta}}{2} .
\label{5_1_30}
\end{equation}
This means that the energy density may be chosen positive.
However, we cannot know beforehand whether in such a case there exists a global, asymptotically flat solution.

In turn, the expressions for the pressures are as follows
\begin{eqnarray}
	\left( p_\chi \right)_0 &=& \frac{e^{-G_0}}{2 \varkappa} \left\{
		\left( F_{, \beta, \beta}\right)_0 +
		\frac{\left( F_{, \beta}^2\right)_0}{2} -
		\left(F_{, \beta}\right)_0 \biggl[
			\left(\Omega_{, \beta}\right)_0 - \left( G_{, \beta}\right)_0
		\biggl]
	\right\} - \frac{e^{3 \cos \beta}}{2 \varkappa},
\label{5_1_40} \\
	\left( p_\beta \right)_0 &=& \epsilon_0 +
	2 \frac{\left( p_{\chi , \beta} \right)_0}{F_{, \beta}} +
	\frac{e^{-G_0}}{2 \varkappa} \left\{
		- \frac{\left(F_{,\beta, \beta}\right)_0\left(\Omega_{,\beta}\right)_0}
		{\left(F_{, \beta}\right)_0} +
		\frac{\left(\Omega_{, \beta}\right)^2_0}{2} +
		\frac{\left(\Omega_{, \beta}\right)_0}{2} \biggl[
			- \left(F_{, \beta}\right)_0 + \left(G_{, \beta}\right)_0
		\biggl]
	\right\} ,
\label{5_1_50}\\
	\left( p_\varphi \right)_0 &=& \epsilon_0 +
		2 \frac{\left( p_{\chi , \beta} \right)_0}{F_{, \beta}} +
	\frac{e^{-G_0}}{2 \varkappa} \left\{
	\left(F_{,\beta, \beta}\right)_0 \left[
			1 - \frac{\left(\Omega_{,\beta}\right)_0}{\left(F_{, \beta}\right)_0}
		\right] +
		\frac{\left(F_{, \beta}\right)_0^2}{2} -
		\frac{\left(F_{, \beta}\right)_0 \left(G_{, \beta}\right)_0}{2} +
		\frac{\left(\Omega_{, \beta}\right)^2_0}{2}
	\right.
\nonumber \\
	&&
	+\left.
			\frac{\left(\Omega_{, \beta}\right)_0}{2} \biggl[
			- 2 \left(F_{, \beta}\right)_0 + \left(G_{, \beta}\right)_0
		\biggl]
	\right\} .
\label{5_1_60}
\end{eqnarray}

\subsection{Particular case $L_{,\beta} = F_{,\beta} = 0$}

Consider now an even more simplified case, when $F_{,\beta}(\chi=0,\beta) = 0$ as well. As one can see from Eq.~\eqref{3_2_20}, this assumes the equality of the tangential pressures at the throat,
\begin{equation}
	\left( p_\beta \right)_0 = \left( p_\varphi \right)_0 .
\label{5_10}
\end{equation}
Next, the pressure $\left( p_\chi \right)_0$ is
\begin{equation}
	\left( p_\chi \right)_0 = \frac{e^{-G_0}}{2 \varkappa} \left[
		\left( \Omega_{, \beta, \beta}\right)_0 +
		\frac{\left( \Omega_{, \beta}^2\right)_0}{2} -
		\frac{\left(G_{, \beta}\right)_0 \left(\Omega_{, \beta}\right)_0}{2}
	\right] .
\label{5_20}
\end{equation}
By comparing this expression to the inequality \eqref{3_2_30}, we have the following inequality for the energy density $\epsilon_0$ and the pressure $\left( p_\chi \right)_0$:
\begin{equation}
	\epsilon_0 + \left( p_\chi \right)_0 < 0 .
\label{5_30}
\end{equation}
According to \eqref{3_2_100}, this assumes the violation
of the null energy condition.
At the same time, the pressures  $\left( p_\beta \right)_0$
and $\left( p_\varphi \right)_0$ remain arbitrary,
obeying only the condition \eqref{5_10}.

\section{Discussion and conclusions}

In order to obtain a toroidal $T^2$ wormhole, it is necessary to solve the following problems:
(a) to assign boundary conditions at the throat;
(b) to assign asymptotic boundary conditions at infinity;
(c) to obtain numerical solutions
(since presumably it will be impossible to get analytical solutions
because of the complexity of the partial differential Einstein equations).

Each of these problems is quite complicated. To solve the problem (a), it is necessary to analyze the energy conditions imposed on the matter supporting the wormhole: whether the violation of the energy conditions is needed or not. Perhaps, the violation of the energy conditions is necessary only
on a part of the torus $T^2$ forming the throat.

To solve the problem (b), it is necessary to find an asymptotic analytical solution of the Einstein equations at infinity. The problem is that the spatial infinity is given by the conditions $\alpha = \beta = 0$ in the metric \eqref{tor_metric}, and the analysis of this (coordinate) singularity is not simple. The behaviour of the metric functions depends on the relation
between $\alpha$ and $\beta$, when they approach zero.

The problem (c) consists in obtaining numerical solutions to the corresponding Einstein-matter equations. These equations will in general be partial differential equations with boundary conditions assigned at the throat and at infinity. It is clear that finding the numerical solution of such a set of equations represents a great challenge.

In the present paper we have studied the problem (a). We have obtained the inequalities describing the conditions needed for a toroidal $T^2$ wormhole to exist assuming that minima of all metric functions (or the area) are reached at $\chi = 0$ for all values of the angular coordinate $\beta$ simultaneously. These conditions are geometric, and they define the requirements for the energy density, pressures, and the metric, providing a minimum of the linear sizes of the cross section of a toroidal wormhole at the throat. Physically, these inequalities describe the energy conditions for the matter supporting a toroidal $T^2$ wormhole. Here we have obtained these conditions in a general form. These conditions have complicated and intricate form; therefore, in order to obtain more concrete results  clarifying the physical situation, we have analyzed the derived  inequalities in some special cases. In one particular case, we have shown that a  $T^2$ throat may exist only when the null energy condition is violated.

\section*{Acknowledgments}
We gratefully acknowledge support provided by Grant No.~BR05236322 in Fundamental Research in Natural Sciences by the Ministry of Education and Science of the Republic of Kazakhstan. We are grateful to the Research Group Linkage Programme of the Alexander von Humboldt Foundation for the support of this research. We also gratefully acknowledge support by the DFG Research Training Group 1620 {\sl Models of Gravity} and the COST Action CA16104
{\sl GWverse}.

\end{document}